\documentclass[sts]{imsart}

\RequirePackage{amsthm,amsmath,amsfonts,amssymb}
\RequirePackage[numbers]{natbib}
\RequirePackage[colorlinks,citecolor=blue,urlcolor=blue]{hyperref}
\RequirePackage{graphicx}
\RequirePackage{booktabs, dcolumn}
\newcolumntype{d}[1]{D..{#1}}

\usepackage[utf8]{inputenc}
\usepackage{xcolor}
\usepackage{wasysym} 
\usepackage{mathtools} 
\usepackage{booktabs}
\usepackage{url}
\usepackage[normalem]{ulem} 

\definecolor{green}{RGB}{0, 153, 0} 
\definecolor{yellow}{RGB}{250, 250, 0} 
\definecolor{orange}{RGB}{214, 129, 0} 
\definecolor{red}{RGB}{179, 0, 0} 
\definecolor{lava}{rgb}{0.81, 0.06, 0.13}
\definecolor{emerald}{RGB}{12, 166, 151}

\newcommand{\dischargedSymbol}{1}
\newcommand{\hospitalizedSymbol}{0}
\newcommand{\ventedSymbol}{2}
\newcommand{\deadSymbol}{3}

\newcommand{\spacedColon}{\mkern .5mu : \mkern 1mu}
\newcommand{\given}{\, | \,}

\startlocaldefs
\theoremstyle{plain}

\theoremstyle{remark}

\endlocaldefs

\begin{document}
	
\begin{frontmatter}
	\title{Learning and Predicting from Dynamic Models for COVID-19 Patient Monitoring}

	\runtitle{Learning and Predicting from Dynamic Models for COVID-19 Patient Monitoring}
		
	\begin{aug}
		\author[A]{\fnms{Zitong} 		\snm{Wang}
		\ead[label=e1]{zwang238@jhmi.edu}},
		\author[B]{\fnms{Mary Grace} \snm{Bowring}%
		\ead[label=e2]{mbowrin1@jhmi.edu}},
		\author[C]{\fnms{Antony}		 \snm{Rosen}%
		\ead[label=e3]{arosen@jhmi.edu}},	
		\author[D]{\fnms{Brian}		 \snm{Garibaldi}%
		\ead[label=e3]{bgariba1@jhmi.edu}},	
		\author[F]{\fnms{Scott} 		\snm{Zeger}%
		\ead[label=e3]{sz@jhu.edu}}	
		\and
		\author[E]{\fnms{Akihiko} 		\snm{Nishimura}%
		\ead[label=e3]{aki.nishimura@jhu.edu}}	
		\address[A]{Zitong Wang is a PhD student, Department of Biostatistics, The Johns Hopkins University Bloomberg School of Public Health, Baltimore, MD, USA.} 
		\address[B]{Mary Grace Bowring is an MD/PhD student, Departments of Biomedical Engineering and Biostatistics, The Johns Hopkins University School of Medicine, Baltimore, MD, USA.} 
		\address[C]{Antony Rosen is Mary Betty Stevens Professor of Medicine and Vice Dean for Research, The Johns Hopkins University School of Medicine, Baltimore, MD, USA .} 
		\address[D]{Brian Garibaldi is Associate Professor, Department of Medicine, The Johns Hopkins University School of Medicine, Baltimore, MD, USA.}
		\address[F]{Scott L. Zeger (co-senior author) is John C. Malone Professor, Department of Biostatistics and Medicine, The Johns Hopkins University Bloomberg School of Public Health, Baltimore, MD, USA.} 
		\address[E]{Akihiko (Aki) Nishimura (co-senior author) is Assistant Professor, Department of Biostatistics, The Johns Hopkins University Bloomberg School of Public Health, Baltimore, MD, USA} 
	\end{aug}
		
	\begin{abstract}
COVID-19 has challenged health systems to learn how to learn. This paper describes the context, methods and challenges for learning to improve COVID-19 care at one academic health center. Challenges to learning include: (1) choosing a right clinical target; (2) designing methods for accurate predictions by borrowing strength from prior patients' experiences; (3) communicating the methodology to clinicians so they understand and trust it; (4) communicating the predictions to the patient at the moment of clinical decision; and (5) continuously evaluating and revising the methods so they adapt to changing patients and clinical demands.

To illustrate these challenges, this paper contrasts two statistical modeling approaches --- prospective longitudinal models in common use and retrospective analogues complementary in the COVID-19 context --- for predicting future biomarker trajectories and major clinical events. The methods are applied to and validated on a cohort of 1,678 patients who were hospitalized with COVID-19 during the early months of the pandemic. We emphasize graphical tools to promote physician learning and inform clinical decision making. 	
\end{abstract}
		
\end{frontmatter}

\section{Introduction}\label{sec:intro}
COVID-19 has challenged health systems around the world. Most infected persons have mild to moderate flu-like symptoms, but a subset have life-threatening lung injury requiring hospitalization and critical care services. Since the first cases early in 2020, more than 5.5 million people have died, constituting the most deadly pandemic since the 1918 Influenza \cite{covid_dashboard}.

The U.S. healthcare system is the most expensive in the world. The excess per capita expenditures relative to the second most expensive country total more than \$1.1 trillion per year or 5\% of U.S. GDP. Despite this excessive investment, U.S. health outcomes are not competitive with countries that spend much less. For example in 2020, an American's life expectancy at age 65 ranked 26th out of 40 reporting OECD countries \cite{lifeexp_65}.  Americans, who comprise 4.2\% of the world population, have suffered 835,000 COVID-19 deaths, 15\% of the global total. Poorly-informed medical decisions, often caused by misaligned incentives, explain a substantial part of the inefficiency in the U.S. health system \cite{berwick2012elim_waste, rosen2019precision}.

One strategy to improve health outcomes and reduce wasteful spending, advocated by the National Academy of Medicine and others, is the {\it learning health care system} \cite{NAP11903}.  
The strategy is to collect the massive amounts of information generated in medical practice, learn from past successes and failures as reflected in these data, and change clinical practice to improve patient outcomes \cite{budrionis2016}. 
During the COVID-19 pandemic, the combination of exponentially-increasing patient numbers, intense life-preserving medical interventions, and poor understanding of how to manage the disease has strained what was already an inefficient system. 

In this paper, we present the statistical models and graphical tools we have developed to support clinical decision processes at the Johns Hopkins Health System in the context of the COVID-19 pandemic.
Also discussed are some of the challenges we have faced in integrating these tools into clinicians' workflow.
Section~\ref{sec:prec_medicine} describes the data and analytic infrastructure available to statisticians and design principles behind our statistical approach in supporting clinical decisions.
Section \ref{sec:introduce_model} introduces the models we developed for COVID-19 patient monitoring. 
In Section \ref{sec:model_application}, we apply the methods to the data set consisting of  the 1,678 patients hospitalized during the early months of the epidemic.
We highlight graphical displays as well as evaluations of the dynamic prediction process starting with the baseline risks and adding regular updates as the biomarkers change. 
Section \ref{sec:choices_challenges} discusses the challenges in implementing our tools and others like them in clinical care and how to assure the tools adapt to the changing patient population and clinical demands. 

"You never want a serious crisis to go to waste," said Rahm Emanuel, White House Chief of Staff for President Obama \cite{wsj2008}. 
COVID-19 is the healthcare crisis that demands that medical systems learn how to learn. 

\section{Precision Medicine, COVID-19, Statistical Framework to Support Clinical Decisions}	
\label{sec:prec_medicine}
At Johns Hopkins, a key component of its strategy to become a learning health care system is its precision medicine initiative called {\it Hopkins inHealth}, where ``in'' is for ``intelligent.''
To enable this initiative, Johns Hopkins a decade earlier installed a system-wide electronic health record (EHR) that collects clinical data and text from all patient encounters. 
In addition to routine clinical data, \textit{Precision Medicine Centers of Excellence (PMCOEs)}  acquire research-grade measurements of their patients.
The Centers collect a range of measurements, including genomic, imaging, physiologic, signs and symptoms, patient reported, and social-behavioral data. 

As an example of this strategy, over the first three months of the pandemic, the COVID-19 PMCOE built the Johns Hopkins CROWN Registry comprising all clinical data related to the testing and treatment of COVID-19 patients including demographic characteristics, medical histories, co-morbid conditions, vital signs, respiratory events, and laboratory values \cite{Garibaldi2021, bowring2021}. During this period, the Johns Hopkins health system admitted 1,978 COVID-19 patients. Of these, 1,687 arrived in the mild/moderate disease state, contributing sufficient clinical data for learning; 1,378 patients were ultimately discharged without ventilation, 199 were ventilated of whom 60 subsequently died, and 110 died without ventilation. 

Learning requires a secure platform for data wrangling and statistical analysis. The EHR data for COVID-19 patients are routinely downloaded into the JH-CROWN registry within a cloud-based system called the {\it Precision Medicine Analytics Platform (PMAP)}. The data are integrated with other external and internal research data and shaped into a clinical cohort dataset organized by patient ID and time. Each PMCOE has a secure platform for analysis using R, Python, Jupyter Notebooks and other standard tools. The clinician scientists in each group are charged with identifying clinically-relevant subgroups and tailoring interventions to improve outcomes.  Hypotheses generated from initial analyses are integrated into decision support tools and tested in clinical settings. 

The reminder of this section describe a statistical framework to support a learning health care system.

\subsection{Choosing a right clinical target}
\label{sec:choosing_clinical_target}
A key clinical objective is to triage patients from highest to lowest risk of requiring invasive interventions. 
Clinicians demand clinically interpretable, dynamic predictive models for the competing risks of intubation and death to focus their resources on patients at greatest risk. In simplest terms, clinicians seek tools that quantify their experiences with past patients as they strive to improve care for similar patients going forward. \\

Consider, from a clinician's perspective, a COVID-19 patient with mild or moderate disease on the day of admission (day 0). The patient presents with personal and medical characteristics $(X)$. The clinician measures baseline levels for a set of biomarkers used to monitor disease progression $(Y_0)$. Examples include pulse rate, oxygen saturation level, and body temperature. At this point, the clinician makes a preliminary assessment of her patient's risk for each of the three competing events: discharge ($\dischargedSymbol$), severe disease defined by the need for ventilation ($\ventedSymbol$), or death ($\deadSymbol$), versus remaining event-free in the hospital for another day ($\hospitalizedSymbol$). We will write $W_{t}$ to denote the patient's status on day $t$, where $W_t \in \{\hospitalizedSymbol, \dischargedSymbol, \ventedSymbol, \deadSymbol\}$. When planning a clinical approach on day $t>0$, the clinician updates her baseline assessment by taking into account the observed trajectories for the biomarkers $Y_{1:t}$.
The clinical target for statistical analysis is the probability distribution of the future competing events and biomarkers given the individual patient's history up until today. Letting $T$ be the time of a person's future event, this conditional distribution can be written
$$[T, W_{T}, Y_{t+1:T}  \hspace{1 mm}  |  \hspace{1 mm} T>t,  \hspace{1 mm} Y_{0:t}, \hspace{1 mm} X].$$

The statistical specialty of {\it joint models} for longitudinal and survival or competing risks data focuses exactly on this joint distribution \cite{chi2006joint, andrinopoulou2014joint}. Joint models are an example of multivariate (many biomarkers and events) hierarchical statistical models that include random effects to represent the heterogeneity among patients in their disease experience \cite{gelman2006data}.   An excellent overview of the methods and computations for joint models is in \cite{rizopoulos2012joint}. A recent expository article on using joint models for predicting non-COVID-19 patient outcomes is by \cite{andrinopoulou2021reflections}. 

We focus on modeling the dynamic biomarkers that signal patient improvement or deterioration. More generally, albeit beyond the scope of this paper, we can account for the influence of past interventions on the biomarkers by modeling both biomarkers and interventions as causally interacting components of the dynamic $Y_t$ process. The reader interested in dynamic treatment regimes is referred to \citet{zhang2018InterpretableDT} and references therein.

\subsection{Borrowing strength from prior patients' experiences} 
\label{sec:method_to_borrow_strength}
A well-designed hierarchical model of the distribution of the future events and biomarkers given the observed history of an at-risk patient is one way to summarize the major sources of variation in prior patient experiences in order to predict what the next otherwise-similar patient is likely to experience. John Tukey referred to this process as {\it borrowing strength} from the prior experience \cite{tukey1963borrowing}. 

 There are multiple complementary ways to decompose the distribution of interest into components that can be estimated from past patient data.  Prospective and retrospective decompositions of  dynamic statistical models are the focus of the remainder of this paper in order to illustrate the challenges to using statistical models to learn to improve COVID-19 care. By using different decompositions, we can see how sensitive predictions are to the specific choice of method for borrowing-strength. 

The more common prospective approach is to decompose the joint distribution of a next interval value $(W_{t+1}, Y_{t+1})$ given the past into the product of: 1) the distribution of each possible state $W_{t+1}$ at time $t+1$, given the person is at risk $W_t = \hospitalizedSymbol$ and given the history of biomarkers $Y_{0:t+1}  \hspace{1 mm} $ and the baseline values $X$, and 2) the probability of the next biomarker value given the biomarker history and baseline measures. This decomposition, that we refer to as a {\it prospective model}, can be written
	\begin{align}
		\label{eq:prosp_decomp}
	\begin{split}
		& [W_{t+1}, Y_{t+1} \hspace{1 mm} | T>t, \hspace{1 mm} Y_{0:t}, \hspace{1 mm} X]  = \\ 
		& [W_{t+1}  \hspace{1 mm} |  \hspace{1 mm} T>t,  \hspace{1 mm} Y_{0:t+1},  \hspace{1 mm} X] \hspace{2 mm} 
		[ Y_{t+1}  \hspace{1 mm} |  \hspace{1 mm} T>t, \hspace{1 mm} Y_{0:t},  \hspace{1 mm} X],
	\end{split}
	\end{align} 
where we use the brackets to denote the (conditional) distributions of random variables.

Focusing on the first term on the right hand side, there are many ``static'' methods that predict clinical events using baseline measures but few that use the history of biomarkers over time. Examples of static prediction models that based upon baseline covariates using logistic regression and/or machine learning are in \cite{Wollenstein-Betech2020,Wang2021Jmed,Bennett2021,Pourhomayoun2021,dabbah2021machine}. Patient features commonly used in these static models include demographic variables (age, sex, race), medical history variables (body mass index, comorbidities), baseline vital signs, and laboratory measurements. A survey of predictor variables used to predict disease severity is in \citet{gallo2021predictors}. These static models are useful to triage patients at baseline but not for patient monitoring day to day as is the goal here. 
Less common are COVID-19 dynamic prediction models using not only baseline predictors but also measured biomarkers that change over time $(Y_{1:t+1})$. 
The machine learning models developed by \citet{chen2021predictive} and \citet{Wongvibulsin2021_scarp} incorporate the dynamic nature of biomakers;
for our purpose, however, key shortcomings of their methods are that both are prospective in design and neither attempts to forecast the biomarkers as in the second term on the right hand side of Equation \ref{eq:prosp_decomp}.

The prospective approach has a few major short-comings in the context of COVID-19. First, the risks of discharge, severe disease and death may be complex functions of the history of current and past biomarkers. Finding a parsimonious set of functions of biomarker values to include in the prediction model is a non-trivial task. Second, the clinician wants to know the risk in both the near and more distant future. Clinicians tell us that their focus is often on predicting the worst outcome in the future, not just on tomorrow's or another day's value. 
Third, day $t=0$ refers to the time of admission not the time of disease onset. By day $0$, each patient has traveled a different disease path introducing substantial baseline variability. 
		
An alternative decomposition that partially addresses these limitations, while introducing others, is what we will refer to as a {\it retrospective model} that uses the day of event $T$ to align patients' disease trajectories rather than the day of hospitalization, as follows: 

	\begin{equation}
	\begin{aligned}
		[T, &W_{T}, Y_{1 \spacedColon T}  \given T > t, Y_{0 \spacedColon t}, X] = \\
		& [Y_{t + 1 \spacedColon T} \given  T, T > t, W_{T}, Y_{0 \spacedColon t}, X] [T, W_T \given T > t, Y_{0 \spacedColon t}, X] \nonumber.
	\end{aligned}
\end{equation}
 
We can then make predictions based upon the following further decomposition:

	\begin{equation}
			\label{eq:aki_decomp}
	\begin{aligned}
		[T, &W_{T}, Y_{t+1 \spacedColon T}  \given T > t, Y_{0 \spacedColon t}, X] \propto  \\
		&[Y_{t + 1 \spacedColon T} \given  T, T > t, W_{T}, Y_{0 \spacedColon t}, X] \\
		&[T, W_T \given T > t, Y_{0}, X][Y_{1:t} \given W_{T}, T, T > t, Y_{0}, X].
	\end{aligned}
\end{equation}
When, after the appropriate transformation, the conditional distribution of $Y_{1 \spacedColon T}$ is reasonably approximated by a Gaussian distribution, it is clear how to compute the first term on the right hand side in \eqref{eq:aki_decomp}. The second term is also easily obtained as shown in the {Appendix} \ref{appn:dynam_prediction}.
	
In words, the right hand side shows that to predict future biomarkers and events, we must estimate three terms:  1) the probability distribution of the future biomarker process given the event type $W_{T}$ and day $T$ and given the observed biomarker process to date $Y_{1:t}$; 2) the probability of each event on each future day given only the baseline biomarkers and patient characteristics; and 3) the probability distribution of the observed biomarker history $Y_{1:t}$ given that the event is of type $W_{T}$ on event day $T$. 
In the {Appendix} \ref{appn:dynam_prediction}, we justify the decomposition using standard conditioning arguments.

As detailed in the sections that follow, we can provide the clinician with both graphical and numerical displays of her patient's future risks and update them with each additional day of biomarker observation from either the prospective or retrospective models. In the retrospective model, we show how implementation is relatively easy because estimating the first term can be done with standard discrete-time competing risks methods. 
And the second term corresponds to a retrospective model of the biomarker data given the event types and times. 
Bowring et al. studied the biomarker distributions just before the onset of the event and their relationships to baseline covariates \cite{bowring2021}. 
The prospective modeling approach is more common, but the retrospective approach has also been beneficial in other related problems which will be discussed in more details in the {Appendix} \ref{appnB:model_retro}.
	

\section{Modeling approaches}
\label{sec:introduce_model}

\subsection{Overview}

We observe repeated values of multiple biomarkers for each patient and seek to predict his/her future biomarker trajectories and event risks from models estimated for a population of patients. 
Hierarchical modeling is a natural choice for this task. 
We use a Bayesian approach to exploit both the computational flexibility of Markov Chain Monte Carlo algorithms and to allow for informative priors where solid prior knowledge exists. 

In the COVID-19 application, we jointly model three continuous biomarkers: pulse rate; body temperature; and SpO$_2$-FiO$_2$ ratio, a measure of lung efficiency calculated by dividing oxygen saturation measurement from pulse oximetry by the fraction of inspired oxygen. 
We assume that the joint distribution of the continuous biomarkers can be reasonably approximated by a linear mixed effects model.  Because it is unlikely that biomarker measurements have marginal distributions well approximated by the Gaussian, we replace the observed biomarker values by the corresponding quantiles of a Gaussian variate. Once models are estimated and predictions made, we transform them back to the original scales. 

In the prospective model for biomarkers’ trajectories, the time origin is the day of admission. 
The multivariate mixed effects model for $[Y_{1:t} \given Y_0, X] $ assumes the mean curve for each biomarker to be a smooth function of time, represented by a natural spline with fixed degrees of freedom.  Baseline covariates $X$ are included as fixed effects. The smooth function is allowed to vary among patients by including the spline bases as random effects. 
This model is used to estimate the mean curves and the covariance of the multivariate response vector for each patient. We then use this model to simulate future realizations for the biomarkers from the distribution $[ Y_{t+1:T} | Y_{1:t}  \hspace{1 mm} , X]$ that are needed in predicting future events. 

To complete the specification of the prospective model, we choose a discrete-time cause-specific hazards model for the three competing outcomes \cite{tutz2016discrete_timetoevent}. 
This model can be fit as a multinomial logistic regression model in which the reference category is remaining hospitalized without an event. The logit of the probability of a particular event (e.g. death) depends on the baseline covariates and on simple functions of each of the biomarkers. In the COVID-19 application, the risk of each event depends on the most recent observation of the biomarkers and on their linear trend over the past 3 days. The multinomial model predicts daily risks of discharge, ventilation, and death as a function of the biomarker histories. 

In the retrospective model, the time origin is the day of the event rather than the day of admission. A potential benefit of this choice is to reduce some of the heterogeneity among patients at the time of hospital admission; e.g. due to different health conditions, viral status, and access to care. As they approach a common outcome, they are likely more homogeneous. The key idea of the retrospective approach is to stratify the patients by their outcomes and build separate multivariate linear mixed effects models for each stratum.
The biomarker trajectories are modeled backward in time, conditional on the type and future date of the outcome. 

As in the prospective model, we use multinomial logistic regression for the distribution of the three competing clinical events: discharge, ventilation, and death. 
The predictors, however, consist only of the patients' baseline characteristics $(Y_0, X)$, not the dynamic biomarker values $(Y_{1:t})$.  For each unique set of baseline predictors, this model gives the estimated probabilities of the competing events for each day since admission. 
We can think of this discrete distribution as the clinician's best assessment of risks at admission but prior to monitoring the patient's biomarker trajectories. 
This initial risk estimates combine with the retrospective multivariate mixed effects model to yield the updated predictive distribution of interest $ [T, W_T, Y_{t+1:T} \hspace{1 mm} |  \hspace{1 mm} T>t, Y_{0:t},  \hspace{1 mm} X].$

The notation and formal model specifications are provided in full detail in the {Appendix} \ref{appnB:model_notation}.

\subsection{Model validation }\label{sec:model_eval_method}
With each additional day of hospitalization, more biomarker data are collected and the at-risk population changes, calling for updated predictions. We hence compare the predictive performance of the prospective and retrospective models on hospitalization days $t = 0, 2, 4 \mbox{ and } 8$. For each model and $t$, we predict both future biomarkers and event probabilities from the next day through to day 20.  
We compare the predicted future biomarker values and event risks with the observed values to quantify prediction error. To estimate out-of-sample prediction errors, we fit each model to a random 80\% of the data and evaluate its prediction error for the remaining 20\%. 
To stabilize the prediction error estimates, we repeat this evaluation for all five 20\% subsets. The $20-80$ split is arbitrary and others might be used. 

Validation of the event risk predictions has two parts: calibration to assess prediction bias by asking whether a predicted rate is close to the observed rate for a new population; and discrimination to assess whether persons with higher predicted risks have events more often than those with lower risks.  For calibration, we contrast the observed and model-based expected numbers of incident events on each future day until day 20. We split the future days into three to five bins for which the expected number of events exceeds five and then calculate a cross-validated chi-square statistic as a measure of departure. 
To assess discrimination, we calculate time-varying area-under-the-curve (AUC) \cite{heagerty2000_timeROC}. 
In particular, we consider the task of predicting the binary outcome of discharge vs.\ severe disease (ventilation or death). 
We take the observed data up to day $t = 0, 2, 4, \text{ and } 8$ and calculate the cumulative probabilities of each competing event at each future day until day 20.
We then use these cumulative probabilities to calculate the time-varying AUC. 

\section{Application and Results}
\label{sec:model_application}
\subsection{Multivariate linear mixed model}\label{sec:application_joint_model}
\begin{figure*}[htb]  
	\centering
	\includegraphics[width = .9\linewidth]{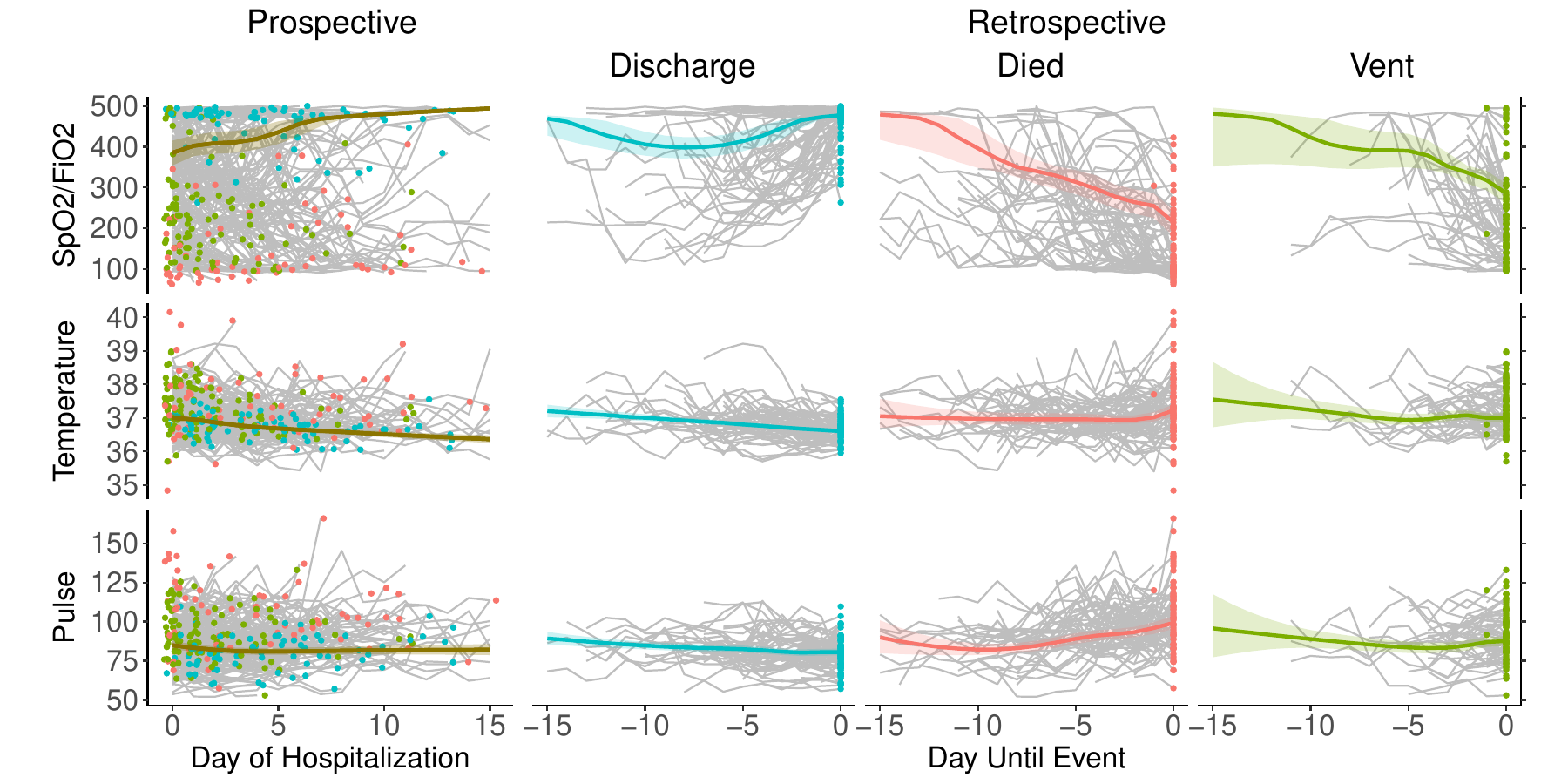}
	\caption{Predicted population-average curves and observed individual patient-level biomarker trajectories from the prospective (left) and retrospective (right) multivariate mixed effects models for three biomarkers. Overall predicted population average curves from the prospective model are shown in dark gold with 95\% credible intervals. Observed patient-level trajectories shown in gray with each outcome indicated by a colored marker terminating the trajectory (blue=discharge, red=death, green = ventilation). Predicted population-average curves from the retrospective event-stratified model are shown in blue, red, and green for discharge, death, and ventilation with 95\% credible intervals. Observed patient trajectories shown in gray with each outcome indicated by a colored marker aligned on day 0.}
	\label{fig:jtgaussian_result}
\end{figure*}

We report on prospective and retrospective models using three key biomarkers: pulse rate, body temperature, and SpO$_2$-FiO$_2$ ratio. In the prospective model, the design matrix ${X}_i$ comprises time (natural spline with 4 degrees of freedom) plus baseline variables including biomarker values at admission, comorbidities and demographics. In the retrospective model, the design matrix $\tilde{X}_i$ comprises three separate time functions, one for each outcome event.
Figure \ref{fig:jtgaussian_result} contrasts the alignment of the biomarker data based on admission time in the prospective model and clinical event in the retrospective model. Biomarker trajectories of a random sample of 100 patients from each outcome group are shown in gray with colored dots indicating the event type and day. The mean curves with confidence regions are model-based estimates and point-wise $95\%$ credible intervals. They represent the mean across the entire sample as if each person were observed at all times.  In the prospective panel, the population average biomarker curve is shown in dark gold for a hypothetical reference patient having population-average baseline biomarker values and reference category characteristics. 
For this reference patient, SpO$_2$-FiO$_2$ ratio increased steadily after admission, body temperature decreased slightly from 37 and pulse stayed relatively constant throughout hospitalization. 
The variation in patient-specific SpO$_2$-FiO$_2$ ratio trajectories is greater than in temperature and pulse trajectories.

In the retrospective panels, biomarker trajectories from admission until event day are shown separately for discharged, ventilated and deceased patients. 
Similar to trajectories observed by \citet{bowring2021}, among those discharged, SpO$_2$-FiO$_2$ ratio increases rapidly starting approximately five days prior to discharge, pulse decreases during hospitalization with a notable drop three to four days immediately prior to discharge, and temperature begins to decrease approximately 10 days prior to discharge. 
Among those who received mechanical ventilation, SpO$_2$-FiO$_2$ ratio decreases five days prior, pulse increases slightly five days prior, and temperature remains elevated prior to ventilation. 
Among those who died without ventilation, SpO$_2$-FiO$_2$ ratio decreases consistently during admission and prior to death, pulse begins to increase approximately 10 days prior to death, and temperature remains elevated during admission and increases immediately prior to death. 
This retrospective approach can be expanded to jointly model more longitudinal biomarkers starting at admission and to add a second phase of follow-up after ventilation. 

For all of the multivariate linear mixed effects models, as a part of routine model checking, we plotted the standardized residuals versus predicted values to identify any systematic deviations from a residual mean of zero (not shown). No systematic deviations were found. Model-based and empirical correlations were compared to within and across biomarkers to validate the modeling assumptions about covariances (not shown). Small systematic departures relative to the size of the correlations were observed, but nothing of scientific interest to warrant changing the model. 
To check the Gaussian assumption for residuals, we produced quantile-quantile plots for the standardized, decorrelated residuals against the standard Gaussian distribution. The Gaussian assumptions were also reasonable.
\\

\subsection{Competing risks model}\label{sec:application_comp_risks_model}

\begin{figure}[p]  
	\includegraphics[width=1\linewidth]{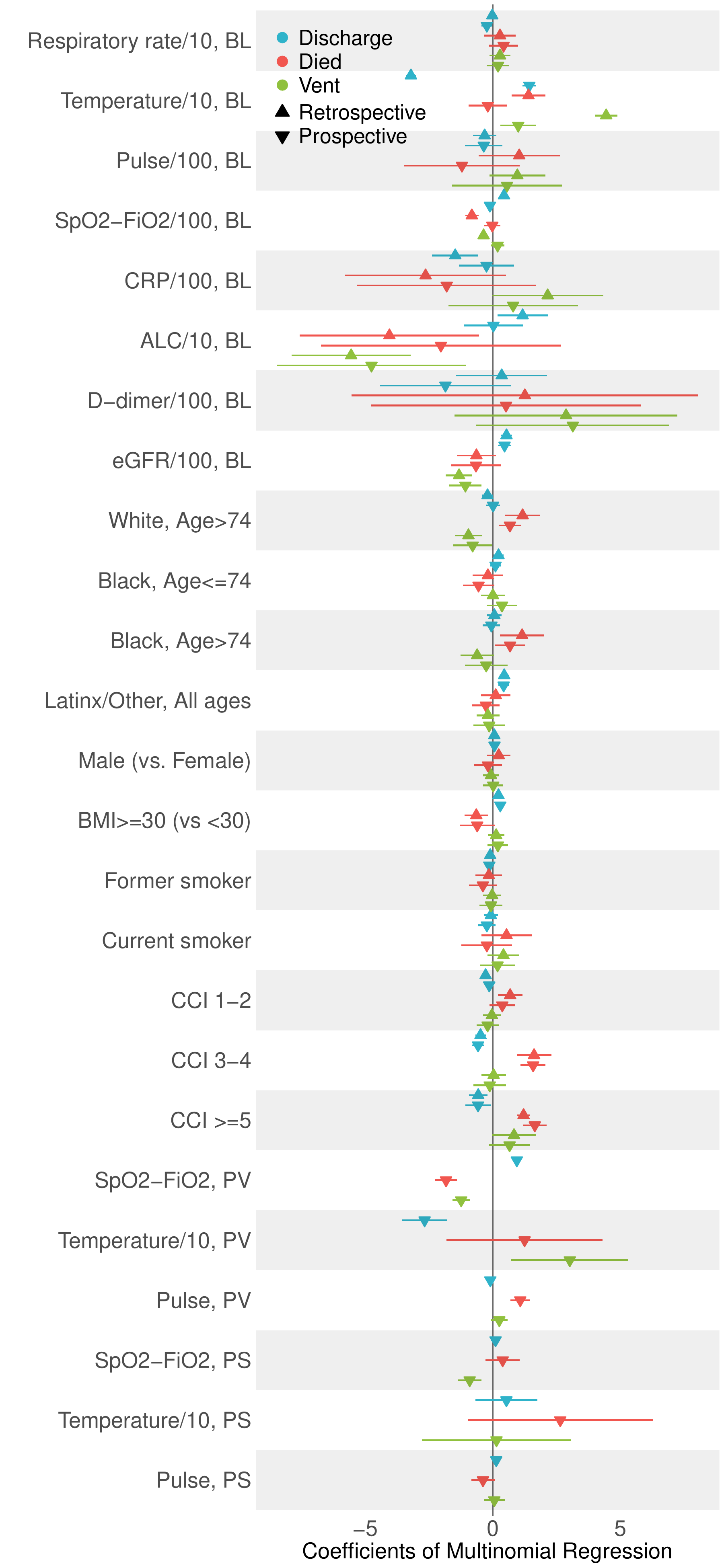}
	\caption{%
		Coefficients and 95\% confidence intervals from multinomial regression under the prospective and retrospective model. Colors indicate clinical events of interest (blue=discharge, red=death, green = ventilation). Up arrows and down arrows indicate the estimates from retrospective and prospective models respectively. Covariates include baseline biomarker values, demographics (sex, age, race), and comorbidities (BMI, smoking status, CCI). 
		The covariates for the prospective model additionally include biomarker values on the previous day and their slopes over the previous two days.
		The following abbreviations are used in figure labels: ALC, absolute lymphocyte count; BL, baseline; BMI, body mass index; CCI, Charlson comorbidity index; CRP, C-reactive protein;  eGFR, estimated glomerular filtration rate; PS, previous slope; PV, previous value.%
	}
	\label{fig:figmult}     
\end{figure}

The competing risks models in both approaches uses the same set of baseline covariates as in the multivariate linear mixed model.
In the the prospective model, we additionally include the biomarker measures on the previous day and their slopes over the previous two days. The estimated regression coefficients with 95\% credible intervals are displayed in Figure \ref{fig:figmult}. 
The combination of the two modeling approaches (prospective and retrospective) and the three outcomes (discharge, death, and ventilation) results in 6 coefficients for each predictor. The figure identifies the important predictors for each outcome and qualitatively compares the results of the two models. Because the prospective model conditions on functions of the past biomarkers and the retrospective model does not, the coefficients are not quantitatively equivalent.

Figure \ref{fig:figmult} shows that, in the prospective competing risks model, increased temperature at baseline is associated with higher log-odds of discharge and increased ALC is associated with decreased log-odds of ventilation. 
When compared to white patients less than age 75, those with Latinx ethnicity have higher log-odds of discharge, Black and white patients over 75 higher log-odds of death, and white patients over 75 lower log-odds of receiving ventilation. 
BMI above 30 is associated with higher log-odds of discharge.
Higher Charlson comorbidity index (CCI) is associated with lower log-odds of discharge and higher log-odds of death. 
Higher previous SpO$_2$-FiO$_2$ ratio, lower previous temperature and lower previous pulse rate are associated with higher log-odds of discharge and lower log-odds of ventilation and death.
An increasing slope in SpO$_2$-FiO$_2$ ratio is associated with a decreased log-odds of ventilation.

In the retrospective competing risks model, most baseline variables including pulse rate, SpO$_2$-FiO$_2$ ratio, CRP, D-dimer, and eGFR have minimal effect on the probability of discharge, death, or ventilation. 
Higher baseline temperature and lower ALC are associated with higher log odds of ventilation and death.
Black patients with age less than 75 and Latinx/Other patients have a higher log-odds of discharge when compared to white patients less than 75. 
Black and white patients over the age of 75 are more likely to die when compared to white patients less than 75. 
Regarding comorbidities, a higher CCI and lower BMI are associated with lower log-odds of discharge and higher log-odds of death.

\begin{figure}[p]
	\centering
	\includegraphics[width=1\linewidth]{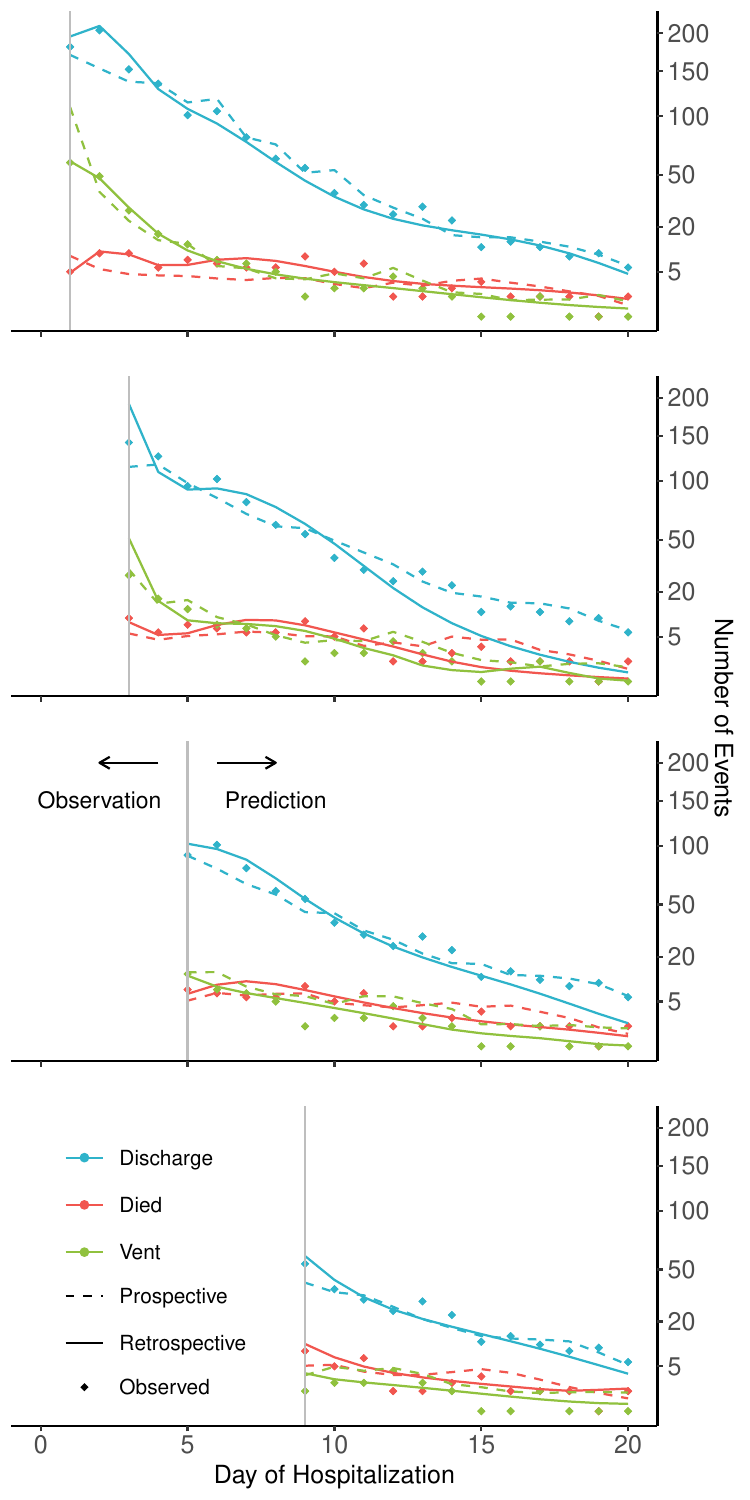}
	\caption{Model calibration. Calibrations of the prospective and retrospective models are illustrated by plotting observed and expected number of events from each model. Each row shows calibration of the models based on dynamically available data ranging from the first 0 day (baseline), 2 days, 4 days and 8 days of hospitalizations for patients still at risk. Colored markers represent the observed number of events at each future time (blue=discharge, red=death, green = ventilation). Solid and dashed lines show the expected number of events from the retrospective and prospective models respectively. Grey vertical lines with arrows separate the observation and prediction periods.}
	\label{fig:calib_discrim_result}
\end{figure}

\subsection{Model evaluation}\label{sec:model_eval}

Figure \ref{fig:calib_discrim_result} evaluates the two approaches by comparing five-fold cross-validated calibration as detailed in Section \ref{sec:model_eval_method}. 
The rows correspond to the at-risk population on days 0, 2, 4 and 8. Grey vertical lines separate observation and prediction periods. 
The colored markers represent the observed number of events ($y$-axis) of each event type (indicated by color) on each future day ($x$-axis). 
Solid lines correspond to the predicted number from the retrospective approach and dashed lines from the prospective approach. 
When using two days of hospitalization data (Figure \ref{fig:calib_discrim_result}, second row), the retrospective approach underestimates the number of patients discharged after day 10. This underestimation translates to an elevated normalized $\chi^2$ as seen in Table \ref{tbl:chisq_auc_table}. Overall, Figure \ref{fig:calib_discrim_result} and the $\chi^2$ column in Table \ref{tbl:chisq_auc_table} show comparable calibration between the two modeling approaches.

We compare the time-varying AUC curves of the two approaches for each remaining at-risk population on days $t=$ 0, 2, 4, and 8. 
The daily AUCs from the two approaches are stable over time and remain comparable, so we only report the AUC value at day 20.
The AUC columns in Table \ref{tbl:chisq_auc_table} show competitive discrimination between discharge and severe disease events. 

\begin{table}
	\tabcolsep=0pt
	\caption{Normalized $\chi^2$ to evaluate calibration and AUCs at day 20 to evaluate discrimination. We calculated the $\chi^2$ statistic for each model and time interval of available data normalized by the  $\chi^2$ degrees of freedom. 
	The AUCs are calculated as described in Section~\ref{sec:model_eval_method}, based on the models' abilities to predict the binary outcome of discharge vs.\ severe disease (ventilation or death).}
	\label{tbl:chisq_auc_table}
	\begin{tabular*}{\columnwidth}{@{\extracolsep{\fill}}lcrcrrr@{}}
		\hline
		&\multicolumn{2}{c}{Normalized $\chi^2$}& & \multicolumn{2}{c}{AUC} \\
		\cline{2-3} 
		\cline{5-6}
		Day &
		\multicolumn{1}{c}{Prospective} &
		\multicolumn{1}{c}{Retrospective}&
		&
		\multicolumn{1}{c}{Prospective} &
		\multicolumn{1}{c}{Retrospective}\\
		\hline
		{0} & 14.38 & 1.82 & &0.83& 0.85 \\
		{2} & 8.72 & 33.08 & &0.83& 0.84 \\
		{4} & 10.34 & 6.99 & &0.85& 0.87 \\
		{8} & 7.80 & 4.02 & &0.87& 0.86 \\
		\hline
	\end{tabular*}
\end{table}

\subsection{Dynamic prediction for individual patient}\label{sec:dynam_pred}

\begin{figure}[b]
	\centering
	\includegraphics[width = 1\linewidth]{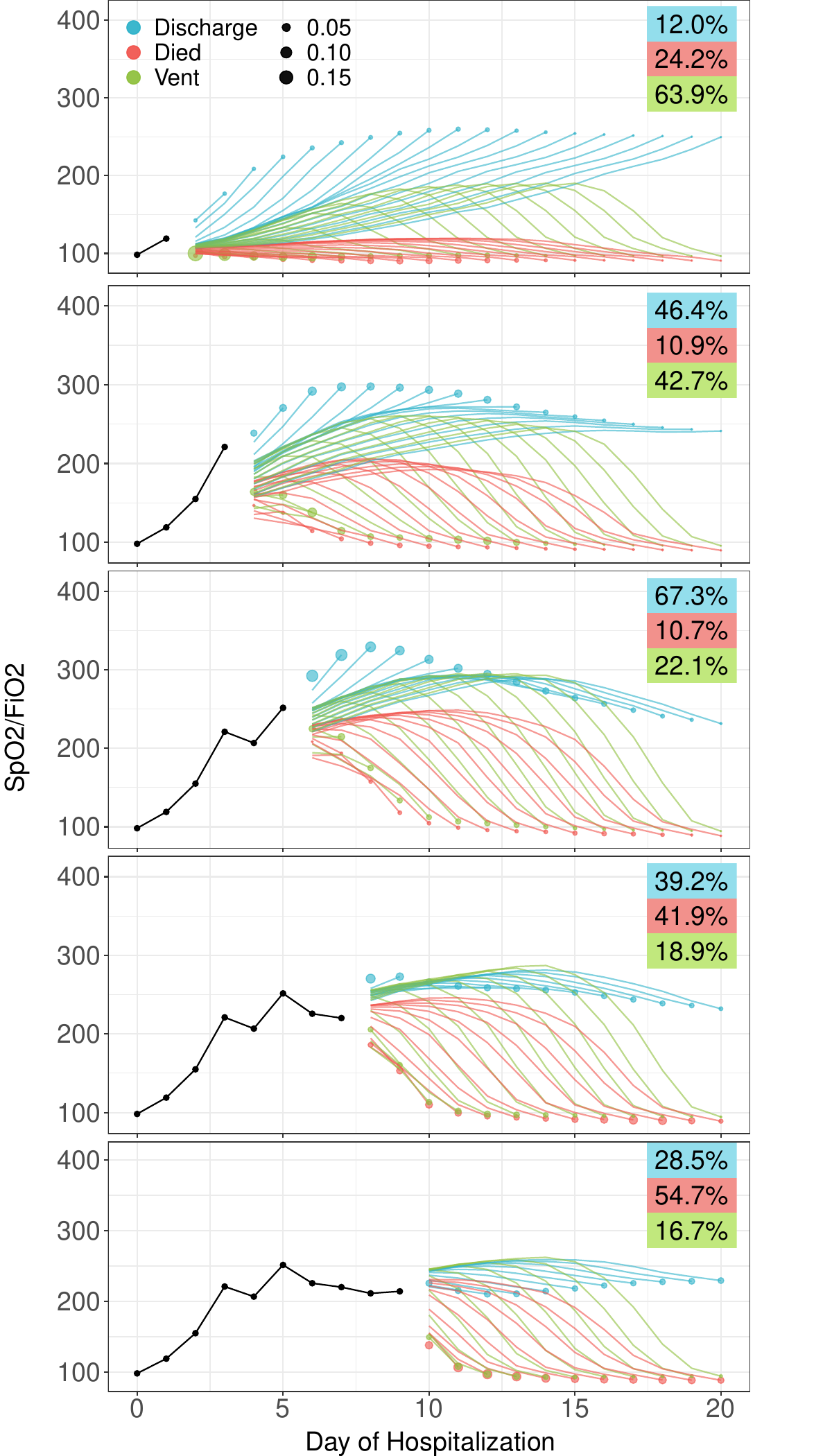}
	\caption{Prediction capabilities of the retrospective model based on data from the first 2, 4, 6, 8 and 10 days of hospitalization for a given patient. Black lines represent the observed SpO2/FiO2 for the patient during hospitalization. Blue, red, and green lines indicate the predicted SpO2/FiO2 trajectories leading to discharge, death, or ventilation events respectively. The marker size at the end of each predicted trajectory indicates the probability of that event on that day given the past biomarker history. The probabilities of each event happening eventually (before day 20) are shown in the colored boxes on the right, with colors indicating event types.}
	\label{fig:figresult}
\end{figure}

Our goal is to provide clinicians with the real-time dynamic predictions for an individual patient of their future risks of the major events and expected biomarker trajectories. 
To illustrate the utility of the retrospective approach in achieving this goal, we randomly select a patient who was discharged on her 13th day of hospitalization. 
This female patient is greater than 75 years old, a former smoker with few co-morbidities and BMI $\geq$ 30. 

The top row of Figure \ref{fig:figresult} shows predicted biomarker trajectories, one for each type of outcome and for each day of the event (up to day 20) after two days of hospitalization. The black line represents the observed SpO$_2$-FiO$_2$ ratio during the first two days. The marker at the end of each trajectory curve indicates the probability of that event occurring on that day given the first two days of observation. For example, if this patient were to be discharged (blue) on day 10, we would expect their SpO$_2$-FiO$_2$ ratio to rise until above 250, their temperature to decrease to 36.4 and pulse to stabilize around 72. However, we can see from the sizes of the event probability markers that this event is unlikely. Finally, the cumulative probabilities (i.e. the sum of the daily probabilities) of events are shown on the right. For this patient, given the first two days of data, the probabilities of discharge, death or ventilation are 63.9\%, 24.2\% and 12.0\%, respectively.

The subsequent rows of Figure \ref{fig:figresult} show the corresponding predictions for this patient, updated using additional days of observation.
For example, after 10 days of hospitalization, the distribution indicates a higher probability of being deceased in the next few days. From the predicted trajectories, if the patient were to be discharged on the next day (day 12), we would expect their SpO$_2$-FiO$_2$ ratio to stabilize over 200. This patient's cumulative probabilities of discharge, death, and ventilation are 28.5\%, 54.7\% and 16.7\%, respectively.

\section{Choices and Challenges}\label{sec:choices_challenges}

\subsection{One or more methods}\label{sec:one_more_methods}
This paper describes what were the second and third approaches to dynamic prediction of COVID-19 outcomes in the CROWN registry. 
It compares two methods derived from different decompositions and modeling assumptions about the same conditional distribution. It is therefore not too surprising that their prediction performances are similar.  
Our first approach applied an extension of random forests for survival models with time-varying predictors RF-SLAM \cite{wongvibulsin2020clinical}, and predicted the occurrence of ventilation or death in the next one or seven days \cite{Wongvibulsin2021_scarp}. 
The statistical models proposed here are comparable to the machine learning approach in terms of the overall accuracy and precision. For example, the one-day ahead prediction cross-validated AUCs from our prospective, retrospective, and their averages are $0.87, 0.86, \mbox{and } 0.87$ respectively, comparable to $0.89$ reported in \cite{Wongvibulsin2021_scarp}. 

The machine learning method is optimized to minimize the prediction error. The statistical models have the advantage that they produce all future predictions for both the events and for the biomarkers. They have more utilities for educating clinicians about what future outcomes to expect. There is a growing realization that, given the complex structure of data generated in health care settings, a statistical method that works well in one situation may not work well in another \cite{madigan2014systematic}. It is essential, therefore, to have alternative models whose predictions can be compared against one another. Moreover, having multiple models can potentially improve over individual models by combining them through model averaging procedures \cite{bates1969combination, hoeting1999bayesian}. 

\subsection{Implementing clinical decision support software}

In our experience, clinical colleagues understand the principle that John Tukey called ``borrowing strength,'' the idea that the experience with past patients can inform a clinician's assessment of risk and her decision about a current patient.  The statistical models described here and the machine learning alternative are complex calculations on data from a population of prior patients to quantify the risk for the patient at hand. A key challenge for data scientists is to convince their clinician colleagues validity of their predictions \cite{shortliffe2018clinical}.

At our own institution, there are standardized policies and procedures for validating novel medical devices but not yet for novel decision support tools.  Publication of the methodology is a necessary but insufficient first step. To obtain buy-in from clinicians, they need to be a part of the design, implementation and testing team. Half of the authors of this paper are clinicians or clinical trainees. The work is also part of a larger COVID-19 clinical/data science project in which a third of the forty members are clinicians. 
During their development, the prospective and retrospective approaches and the machine learning alternative were discussed at weekly meetings of the larger team over many months. The tools were more readily accepted given the clinicians' role in their development.

Another key component of validation is the ability of the clinician users to "look under the hood" so as to understand where the predictions come from in general terms \cite{shortliffe2018clinical}. Figure \ref{fig:figresult} is an example of an effort to provide clinicians more than just prediction probabilities. It shows both the event probabilities and the expected biomarker trajectories for each of the clinical events: discharge, ventilation and death. Clinicians can compare these quantitative findings against their qualitative experiences having cared from many patients. For example, it makes sense to clinicians that the SpO$_2$-FiO$_2$ ratio needs to rise into the 250-350 range before a patient can be discharged. Clinicians also want to "kick the tires" of the method by asking "what if" questions. How does the prediction for this patient change if we restrict the training sample to patients 70 years of age? Which prior patients are most like this one and what happened to each of them? A task we have yet to complete is building a Shiny interface that facilitates "what if" questions elicited from clinical colleagues.  

Finally and most importantly, clinical decision support software is regulated within the U.S. by the FDA \cite{karnik2014fda,yaeger2019united}, under its 2019 draft guidance \cite{fda2019cds}. To gain regular use in patients, software to implement the methods described here must undergo clinical evaluation and receive FDA approval. Most institutions currently leave the process of seeking FDA approval to individual research groups to conduct the requisite experiments and file for approval. Lack of an institution-wide regulatory process is a large disincentive to research groups. The more common course for researchers is to form a company under their university's technology transfer policies, raise money for clinical testing, then the company seeks FDA approval. The company can then license the approved software back to the institution where it was developed. This is a slow and arduous process that is a strong impediment to modern data science impacting clinical care. 
At Johns Hopkins, our current strategy is to implement new policies and procedures that expedite seeking regulatory approval for decision support software.

\subsection{Intelligence internal and/or external to the EHR}

The inHealth PMAP system, external to the EHR, is a secure cloud-based system where all of our clinical data analysis must take place to meet federal privacy rules. PMAP routinely downloads then integrates the clinical data with other sources including images, -omics,  social-behavioral, and other measures. This external system was created to wrangle the raw clinical data into clinical cohort datasets, conduct population-level analyses, produce predictions for individual patients, then return the results back to the EHR for use in clinical decision making. The advantage of moving the data out of the EHR is that non-clinical data scientists can collaborate with clinicians to build learning systems using hardware and software necessary for large, complex analyses. These analytic platforms can be scaled separately from the EHR systems. Data can be de-identified and studied with fewer patient-privacy concerns.

The external approach has its drawbacks. 
The first is that new data are not immediately available to external algorithms to generate predictions for a current patient visit. 
Our current approach is to produce two algorithms:  the first uses all of the population data available in PMAP to train the model; the second is a less computationally-expensive updating algorithm that can run within the EHR system to update the predictions based upon this patient's current-visit data. Scaling this two stage approach is difficult because the updating programs must be custom designed for each data type and prediction algorithm. 

The second issue is that, in order to document the clinical process, the predictions must be presented to clinicians and patients within the EHR-directed workflow. 
A similar issue occurs when clinicians use images in guiding patient care.
External systems store and analyze the images. 
Clinicians view the the images that reside outside of the EHR.
Clinicians then write an EHR-based note to document their findings and decisions. 
Image system vendors build the interfaces for this hand-shaking process. 
No analogous interface currently exists to document the process if clinicians were to exit the EHR, view prediction from an external algorithm, and use this external information to guide their decisions.
Such an interface needs to be developed for decision support tools.

\subsection{Continuous model  re-evaluation}\label{sec:model_curation}

Devices within medical systems are recalibrated on a regular schedule to assure the validity of their measurements.  Repeated re-evaluation, what we refer to as "curation" is just as essential for decision support tools, especially in dynamic pandemics like COVID-19. The virus and the patient populations are continuously changing. A system trained on early patients may not be as useful in predicting outcomes for later patients. The standard methods of evaluation illustrated in this paper are necessary but not sufficient to handle the dynamics. Curation comprises repeated routine re-training and re-evaluation of the models on early, middle and late subsets of the growing population of patients. A graphical curation dashboard might include a sequence of figures like Figure \ref{fig:calib_discrim_result} along with Table \ref{tbl:chisq_auc_table} showing calibration and discrimination in each time period. In addition to routine curation, model version updates will likely be needed to adapt to the changing epidemiology. Updates must be version controlled and documented for clinical oversight. When a decision support tool is used in clinical care, the predictions and clinical decisions to which they contributed must be linked to the specific version of the software and training data from which they were generated so that the predictions are entirely reproducible. Curation also requires an open feedback channel lest clinicians using a tool have concerns about predictions for a particular patient. Each major clinical support tool requires a curation team comprising independent clinical, data and regulatory experts who review the routine updates and respond to clinical feedback in a timely fashion. 

\section{Discussion}\label{sec:discussion}

This paper illustrates one academic health center's efforts to learn how best to monitor COVID-19 patients' disease states and predict major outcomes in near real time. 
The infrastructure necessary to build this learning health system comprises: automated acquisition of clinical data via an electronic health record; a secure analytic platform where raw transactional data can be transformed into research-grade clinical cohort data and integrated with external auxiliary data; statistical models for learning in real time; 
accessible presentations of the new knowledge to clinicians and patients within the EHR-driven workflow; capacity to clinically test the value of the new knowledge in clinical trials; and a system for scaling and curating learned knowledge to assure it improves clinical outcomes and controls costs over the longer-term. 

While discussing many of the above components, this paper focused on using longitudinal statistical models to learn about an individual patient’s biomarker trajectories and outcome risks from a population of otherwise similar patients who came before.  We discussed two complementary approaches to decomposing the dynamic probability distribution of  future biomarkers and events given patient histories. 
The traditional prospective approach models the marginal distribution of the biomarker processes and then the risk of events given the biomarkers. The more novel retrospective approach models the baseline risks of the events and then the conditional likelihood of the biomarkers given the eventual outcome. In our analysis of the 1,678 patients, the discrimination and calibration of major events were qualitatively similar between the two approaches.

One goal of this paper is to illustrate a graphical representation of the joint biomarker-event distributions that is accessible to clinicians and is consistent with how they think. For a clinician concerned mainly with triaging patients using their relative risks over the next 24 or 48 hours, the prospective model has the advantage of directly modeling the competing outcomes risk as an explicit function of the recent biomarker values. However, for clinicians concerned about longer-term outcomes and what the expected biomarker trajectories look like for each event type and future date, the retrospective model is a valuable complementary tool.  

The findings in Figure \ref{fig:calib_discrim_result} are generally consistent with current medical knowledge. 
It is well known that rising SpO$_2$-FiO$_2$ ratio and decreasing pulse portend earlier discharge and reduce the risk of intubation and death. One counter-intuitive finding was that obese persons (body mass index (BMI) > 30) were more likely to be discharged and less likely to die after conditioning on other variables. A similar finding has previously been reported from a separate analysis of the JH CROWN registry by \cite{Garibaldi2021}.
Despite adjustment for age and race, this finding might be driven by the older more frail patients in our cohort, for whom higher BMI might be protective.

This paper focuses on the early learning about COVID-19 using the 1,678 patients hospitalized during the early months of the epidemic. The JH CROWN registry now includes more than 8,000 COVID-19 patients. An important systems-level question relevant to the ultimate impact of these and similar tools is how they are to be curated. 
In purchasing devices and pharmaceutical products, health systems count on their suppliers to monitor the longer-term efficacy and efficiency of their products as the population of patients and medical practice change over time. Currently, tools like the ones in this paper do not come with a dynamic curation process to sustain them over the longer-term. Should data science tools be sold like software, imagers, sequencers or drugs? If so, how should they be locally optimized across diverse populations and over time?

Finally, this paper places traditional statistical research within the context of the COVID-19 epidemic, in which healthcare system requirements have made it challenging to achieve routine, intelligent use of emerging data to improve care.  A clearly documented, useful method that predicts the likely trajectories and risks for a current patient is only an important first step. The ultimate goal is for the learned information to improve clinical decisions and patient outcomes. Improvement requires health system infrastructure that is just starting to be built. As it emerges, statisticians and other data scientists have the opportunity to play essential roles in improving health outcomes at more affordable costs.

	\begin{acks}[Acknowledgments]
		The authors would like to thank Dr. Brian Caffo, the referees, Associate
		Editor and Editor for their constructive comments that improved the
		quality of this paper. We also thank the many persons who built the CROWN registry and the CADRE team that has administered its use.
	\end{acks}
	

\begin{appendix}\label{appn}
	
\section{Details on Retrospective Model dynamic prediction}\label{appn:dynam_prediction}

Appendix A provides details on the decomposition formula \eqref{eq:aki_decomp} introduced in Section~\ref{sec:method_to_borrow_strength} for making predictions from the retrospective model.
We suppress the subscript $i$ for individuals to simplify the notation. 
The goal is to compute the joint distribution of the time of event $T$, type of event $W_T$ and future biomarkers $Y_{t+1 \spacedColon T}$ given the individual remains at risk $T>t$, the biomarkers observed to that point ${Y_{1 \spacedColon t}}$, baseline biomarkers $Y_0$ and patient characteristics $X$, (i.e. $[T, W_T, Y_{t+1\spacedColon T} \given T>t, Y_{1:t}, Y_0, X]$).
Given the baseline biomarker measures $Y_{0}$ and patient characteristics ${X}$, the initial probability distribution for event type and day in the interval $(t+1:T)$ is given by
\begin{equation*} %
	\begin{split}
	&[{T = \tau, W_{T} = m} \given T>t, Y_0,{X}] := \xi_{\tau m} \\
	&= \pi_{\tau m} \textstyle \prod_{s = t + 1}^{\tau - 1} \left( 1 - \pi_{s1} - \ldots - \pi_{sM} \right),
	\end{split} 
\end{equation*}    
where the probabilities $\pi_{sm}$ of event type $m$ on day $s$ have been estimated from the retrospective competing risks model.

The likelihood $L(\tau,m; y) $ of observing the biomarkers ${Y_{1:t}} = y$ given the person is at risk at time $t$ and given the event ${W_{T}}=m$ happens at time $T = \tau$ is
\begin{equation}
	L(\tau,  m ; y) =  [{Y_{1:t} = y} \given T = \tau  > t, W_{T} = m,Y_0, {X}]. \nonumber
\end{equation}
This is the conditional distribution of the biomarker up until time $t$ given the outcome event between times $t$ and $T$ that can be estimated from the fitted retrospective multivariate linear mixed effects model.\\

The updated probability  $[{T, W_{T}} \given T>t, {Y_{1:t}},Y_0, {X}]$ can now be obtained by applying conditional probability rule:
\begin{equation}
	\begin{split}
		&[{T, W_T} \given T > t, {Y_{1:t}}, Y_0, {X}] \\
		& =  \frac{
			[{Y_{1:t}} \given T, T> t, {W_T},  Y_0, X]
			[T, {W_T}\given T>t, Y_0, X]}
		{\sum_{{T, W_T}} [{Y_{1:t}} \given T, T> t, {W_T},  Y_0, X]
			[T, {W_T}\given T>t, Y_0, X]} \\
		& =\frac{L(\tau, m; y)\xi_{\tau m}}{\sum_{ \tau, m} 	L(\tau, m; y)\xi_{\tau m}}. \nonumber
	\end{split}
\end{equation}

Finally, we obtain the joint distribution of the future biomarkers and events given the history of biomarkers and sequence of events up until time $t$ such that
\begin{equation}
	\begin{split}
	 &[{T, W_T}, {Y_{t+1 \spacedColon T}} \given T>t, {Y_{1\spacedColon t}}, Y_0, X]  \\ 
		&= [{Y_{t+1 \spacedColon T}}\given T, {W_T}, {Y_{1:t}}, Y_0, X]
		[{T, W_{t+1:T}} \given {Y_{1:t}},Y_0, X]. \nonumber
		\end{split}
\end{equation}

\section{Model details}\label{appnB:model}

\subsection{Notation}\label{appnB:model_notation}
\begin{figure*}[h]  
	\centering
	\includegraphics[width=0.8\linewidth]{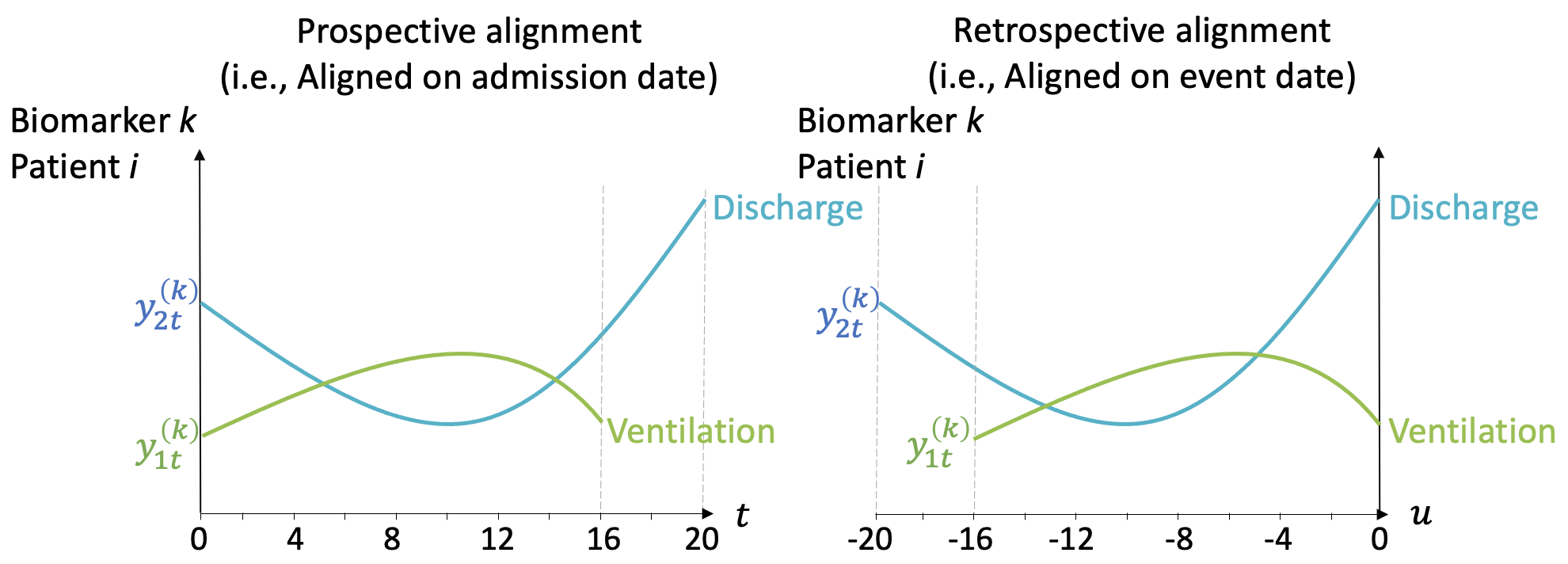}
	\caption{Illustration of follow-up time orientation in the prospective and retrospective models. The prospective model aligns a patient's biomarker data on admission date (left-most axis) and follows them forward in time along the $t$-axis until clinical event. The retrospective model aligns a patient's biomarker data on clinical event date (right-most axis) and follows them backward in time along the $u$-axis until admission date.}
	\label{fig:fignotation}     
\end{figure*}

Appendix B provides model details on the prospective and retrospective approaches. Figure \ref{fig:fignotation} introduces notations for the prospective and retrospective models using two hypothetical patients who were discharged on day 20 and ventilated on day 16, respectively. 
The red vertical lines indicate the time origins for each model; the dashed gray lines indicate each patient's day of event. We use $Y_{it}^{(k)}$ to denote measure $k$ for person $i$ on day $t$ after hospitalization, $i = 1,...,N; t=1,...,T_{i}, k=1,...,K$. 
We let $Y_{i}^{(k)}$ be the vector of $(Y_{it}^{(k)},t = 1,...,T_{i})$. 
Define ${Y_i} = ({Y_i^{(k)}}, k=1,\ldots,K)$.

In the prospective and retrospective multivariate linear mixed effects models, we define the design matrices for person $i$ and biomarker $k$ to be $X_i^{(k)}$ and $\tilde{X}_i^{(k)}$ respectively. Let ${X_i} = \oplus_{k=1}^{K} {X_i}^{(k)}$ and ${\tilde{X}_i} = \oplus_{k=1}^{K} {\tilde{X}_i}^{(k)}$. To reduce notation, we include the baseline biomarker measures $Y_{i0}$ as columns within these design matrices.
Similar naming conventions are used in the competing risk multinomial models, where the prospective and retrospective design matrices are called $G_i$ and $\tilde{G}_i$ respectively.
Finally, in the retrospective model (Figure \ref{fig:fignotation}, right), patients are aligned according to their event date as shown by the red vertical line so we introduce $u_{it} = t - T_{i}$ to represent the day until the event occurred for patient $i$. 

\subsection{Prospective model}\label{appnB:model_prosp}
As mentioned in Section \ref{sec:method_to_borrow_strength}, examples of COVID-19 prediction models using biomarker history over time are limited but do include \citet{chen2021predictive} and \citet{Wongvibulsin2021_scarp}. 
Both methods rely on machine learning algorithms.
The former work combines machine learning methods to select time varying predictor variables with joint models of mortality and biomarkers to produce dynamic mortality predictions. 
\citet{Wongvibulsin2021_scarp} use an extension of random forests for longitudinal, multivariate and survival data to predict the occurrence of severe disease in the next 24 hours or 7 days given the history of dynamic biomarkers to date.  
The joint model in the Chen article is similar to the prospective approach described below, however, it does not take account of the competing risks of discharge, ventilation and death. 
The Wongvibulsin approach is entirely machine learning to predict severe disease.  A separate algorithm is needed for each of two future interval lengths, one and 7 days. 
Neither of the methods forecasts the biomarkers.

\subsubsection{Multivariate linear mixed effects model} 
We use a multivariate linear mixed effects model to estimate the average trajectory of the $K$ biomarkers and the covariance matrix of observations across measures and time. 
We define  ${X^{(k)}_i}$ and ${Z^{(k)}_{i}}$ to be the $(T_{i} \times p_X), (T_{i} \times q)$ known matrices of full rank, where ${Z^{(k)}_{i}}$ is a subset of $X^{(k)}_i$. ${\beta^{(k)}}$ and ${b^{(k)}}$ are $(p_X \times 1)$ and $(q \times 1)$ vectors of parameters for fixed and random effects respectively.

To simplify the exposition, we assume the number of observations is the same for all $K$ measures but this assumption is easily relaxed in practice. We define  ${X^{(k)}_i}$ and ${Z^{(k)}_{i}}$ to be the $(T_{i} \times p_X), (T_{i} \times q)$ known matrices of full rank, where ${Z^{(k)}_{i}}$ is a subset of $X^{(k)}_i$. ${\beta^{(k)}}$ and ${b^{(k)}}$ are $(p_x \times 1)$ and $(q \times 1)$ vector of parameters for fixed and random effects respectively. The model is specified as follows:

	$${Y_i}^{(k)} = {X_i}^{(k)} {\beta}^{(k)}  + {Z_i}^{(k)}  {b_i}^{(k)}  + {\epsilon_i}^{(k)}$$
where the correlations among the biomakers are modeled via the random effect ${b_i} = ({b}_{i}^{(1)'}, \ldots, {b}_{i}^{(K)'})' \stackrel{ind}{\sim} G_{Kq}({0}, {D})$ and via correlated noises ${\epsilon}_{i} = ({\epsilon}_i^{(1)'},..., \\ {\epsilon}^{(K)'})' \stackrel{ind}{\sim} G_{K T_i}({0}, {R_i})$.

In the prospective model, the fixed effect covariates ${X_i}$ include baseline patient characteristics, basis functions (e.g natural splines) representing a smooth function of days since admission for each biomarker, and possibly interactions. 
The random effect covariates ${Z_i}$ include the basis functions for time to allow the biomarker trajectories for an individual to deviate from the mean trajectories. 
The covariance matrix of the random effects and the residual covariance matrix are assumed to be unstructured. In some applications, they might also depend on baseline covariates but that possibility was not pursued here. 
At any time $t$, this prospective model defines the conditional Gaussian distribution of the unobserved future biomarker values ${Y_{i(t+1:T_i)}}$ given the observed values ${Y_{i(1:t)}}$ for patient $i$. Hence, we can simulate from this conditional distribution and obtain a predictive distribution for future biomarker values.  This model can be implemented using multivariate linear mixed effects software such as MCMCglmm, RStan or the equivalent.

\subsubsection{Competing risks model given baseline covariates and past biomarkers} 
We construct the discrete-time cause-specific hazards as a multinomial logistic model for the conditional probability that an at-risk person has event type $m$ on day $t$. 
Let $\pi_{itm}$ be this discrete time hazard, $m = 1, \ldots, M$ where $M = 3$ in our application and $m = 0$ is the reference category of no event at time $t$. 
Let ${f_{it}} = {f(Y_{i1},\ldots,Y_{it-1})}$ be an $r$-vector of known functions of past biomarkers. We then assume the competing risks model has the form
\begin{equation}
\log\left( \frac{ \pi_{itm} }{ \pi_{it0}} \right)  = {g_{it}}{\gamma_{m}} + {f_{it}} {\alpha_m}, \hspace{2mm} m = 1, \ldots, M, \nonumber
\end{equation}
where $g_{it}$ is the $t$th row of $G_{i}$; ${G_{i}}$ is $(T_{i} \times p_G)$ full rank design matrix for each event type $m$ including the baseline covariates and basis functions that define a smooth baseline hazard function; ${\gamma_{m}}$ is a $(p_G \times 1)$ event $m$-specific vector of parameters; and ${\alpha_{m}}$ is a $(r \times 1)$ vector of regression coefficients representing the influence of the biomarkers on the risk of event $m.$  In this application, we use $r=2$ functions of each biomarkers, the most recent value and the linear slope over the past two times. That is,  ${f_{it}}= ({Y_{it-1}}, {{Y_{it-1}} - {Y_{it-2}}})$.

\subsubsection{Prediction}\label{appnB:prosp_prediction}
Given that an individual remains at risk on day $t$ and given the observed biomarker process until day $t$, the clinician needs risk estimates for a range of future days. 
But the competing risks model above depends on as-yet-unobserved biomarkers starting on day $t+1$. To calculate estimates of the discrete hazards on future days, we must integrate over the distribution of the as-yet unobserved predictors given their past observed values. We use Monte Carlo integration by simulating $S$ realizations of the future biomarkers from this conditional Gaussian distribution, calculating the corresponding ${f_{it'}}, t'>t$ values for each simulation, calculating the predicted discrete hazards on future days from the competing risk model using the simulated ${f}$s, and then average these across the all $S$ simulations. To assure that the Monte Carlo integration error is negligible, we compare the predictions  for $S$ and $S/2$ to assure the predictions are equivalent. 

\subsection{Retrospective model}\label{appnB:model_retro}
The prospective modeling approach is more common, but the retrospective approach has also been beneficial in other related problems.  
Sliced inverse regression (SIR) is one example where the goal is to identify and fit a relatively small number of linear combinations of a large number of potential predictors \cite{duanli1991}. SIR corresponds to modeling $X$ given $Y$ rather than $Y$ given $X$. 
Jiang, Yu and Wang extended the SIR approach for application to longitudinal predictors as in our problem \cite{jiangyuwang2014}. 
Their focus is on the theory of this approach in the context of continuous outcome variable, not competing risks. 
Another closely related decomposition is {\it pattern mixture} models used to handle possibly non-ignorable missingness in analyses of longitudinal data \cite{little1993pattern_mixture}. 
In that context, $W$ is the indicator of whether a person dropped out of the study or not. 
The scientific focus is on the relationship of $Y$ with $X$ absent drop-outs. 
To account for possibly informative dropout process,  \citet{little1993pattern_mixture} and \citet{michiels1999selection_model} model the conditional distribution of $Y$ given $X$ and given the observed dropout time $W$.

\subsubsection{Events stratified model}

To retrospectively model biomarker trajectories, we stratify the the event outcome $W_t=m$ and use a $m-$specific multivariate Gaussian mixed-effects model for $[{Y_i^{(k)}}|{\tilde{X}^{(k)}_i}]$ for measure $k$ that can be written as follows,
$${Y_i}^{(k)} = {\tilde{X}_i}^{(k)} {\tilde{\beta}_m}^{(k)}  + {\tilde{Z}_i}^{(k)}  {\tilde{b}_i}^{(k)}  + {\tilde{\epsilon}_i}^{(k)}$$ where $m=1,2,3$.
In the applications below, the predictor variables include a separate set of natural spline basis functions ${B_j(u_i)}, j=1,..df$ of the time until the event $(u)$ and baseline biomarker measures and baseline covariates. The retrospective model is completed by specifying the dependence of the random effects covariance matrix and residual variances on the event outcome $m$. The other assumptions follow the prospective  model specification as summarized above.

\subsubsection{Competing risks model given baseline measures and baseline covariates} 
In the retrospective model, the competing risks depends only on the baseline biomarker measures and covariates, not on subsequent biomarker values as in prospective model. Otherwise the two models take the same form. 

\subsubsection{Prediction} 
%
From the retrospective competing risks model and the multivariate linear mixed effects model, we obtain the distribution of the date and type of event given observed biomarker values for patient $i$ and their baseline covariates, $[T_i, W_{iT_i} \given T>t, Y_{i(1:t)}, \tilde{X}_i]$, using a prediction updating algorithm. Thus, we can calculate  the target distribution $[T_i, W_{iT_i} Y_{i(t+1:T)} \given T>t, Y_{i(1:t)}, \tilde{X}_i]$ for this patient by multiplying the probability distribution above with the conditional Gaussian likelihood of the future biomarker values given the past observations, baseline values and patient characteristics, $[{Y_{i(t+1 \spacedColon T_i)}}\given T_i, {W_{iT_i}}, {Y_{i(1:t)}},  \tilde{X}_i]$. 


An advantage of the prospective model is that it explicitly expresses the risks of events in terms of known functions of the previously observed biomarkers. 
The form of this dependence is indirectly specified in the retrospective model. 
More recent biomarker values are not as emphasized in the retrospective prediction as they are in the prospective prediction. 
In practice, we found that conditioning on a few proximal observations improves calibration and discrimination in the COVID-19 application.
To do this, we let $a < t$ be the dimension of the desired subspace of ${Y_{it'}}, t' \leq t$ and let ${A}$ be an $aK \times T_i K$ matrix so that ${Y^*_i }= {A Y_i}$ is taken as the new data. For a given type $m$ and date $T_i$ of the future event, the conditional mean and variance of ${Y^*_i }$ can be calculated from the fitted retrospective multivariate linear mixed effects model. We combine this local likelihood with the initial probabilities to produce updated probabilities of the type and day of future events.

\end{appendix}

	\begin{funding}
	Dr. Zeger was partially supported by NIH Grants P30AR070254, 5UL1TR003098 and 1U54CA260492 and by the Scleroderma Research Foundation Grant IPN 21054327. Dr. Garibaldi is a member of the FDA Pulmonary-Asthma Drug Advisory Committee; is a consultant for Janssen Research and Development, LLC; has received speaker fees and served on an advisory panel for Gilead; has received speaker fees from Atea. Ms. Bowring is supported by NIH grant T32GM136577.
	Ms.\ Wang is partially supported by the Patrick C. Walsh Prostate Cancer Research Fund.
	
	The COVID-19 PMAP Registry was funded by Johns Hopkins Medicine through the Precision Medicine Program. The studies were also supported through the generosity of the collective community of donors to the Johns Hopkins University School of Medicine for COVID-19 research.
	
	This work was supported by funding from John Hopkins inHealth, the Johns Hopkins Precision Medicine initiative through JH-CROWN, and the coronavirus disease 2019 (COVID-19) Administrative Supplement for the US Department of Health and Human Services (HHS) Region 3 Treatment Center from the Office of the Assistant Secretary for Preparedness and Response.
\end{funding}

\begin{supplement}\label{supp}

\renewcommand{\thesection}{S\arabic{section}}
\renewcommand{\thetable}{S\arabic{table}}
\renewcommand{\thefigure}{S\arabic{figure}}
\renewcommand{\theequation}{S\arabic{equation}}

\setcounter{section}{0}
\setcounter{figure}{0}
\setcounter{table}{0}
\setcounter{equation}{0}

\stitle{Supplementary Figures}

\begin{figure*}[p]  
	\centering
	\includegraphics[width=.9\linewidth]{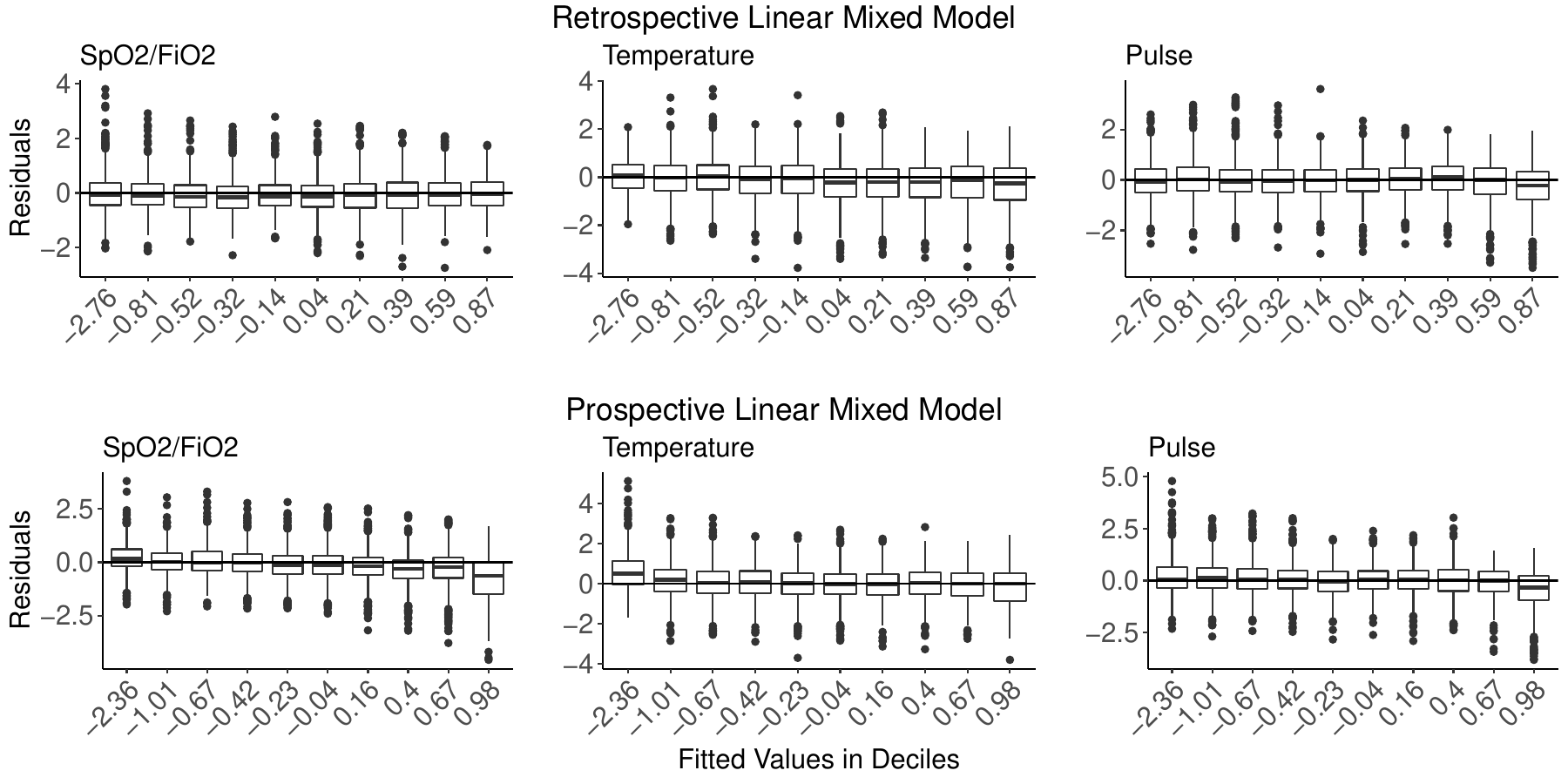}
	\caption{Residuals vs. fitted values in deciles boxplot of prospective (bottom) and retrospective (top) linear mixed effects models for 3 selected biomarkers.}
	\label{fig:resid_fitted_hist}     
\end{figure*}

\begin{figure*}[p]  
	\centering
	\includegraphics[width=.55\linewidth]{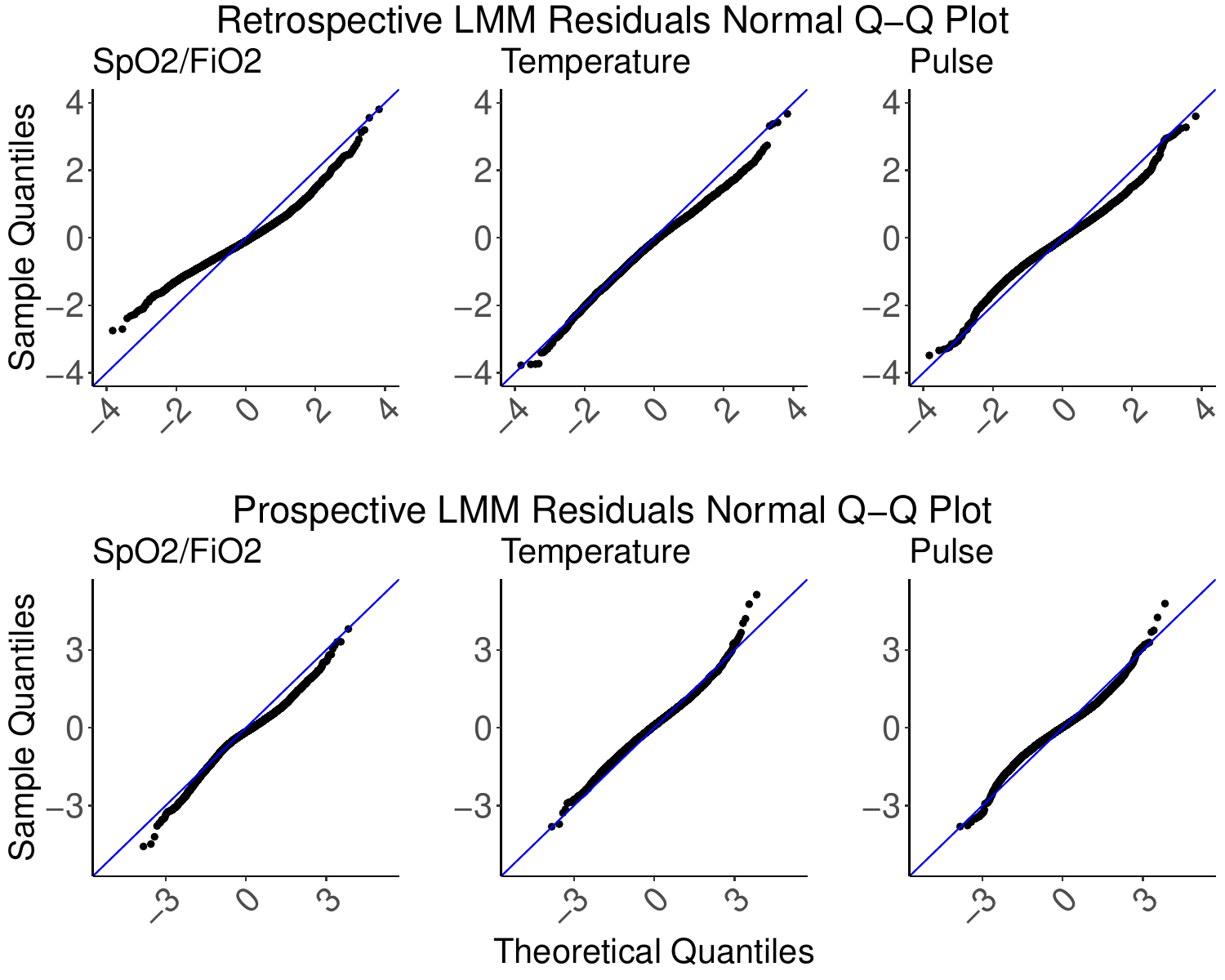}
	\caption{Normal Q-Q plot for residuals of prospective (bottom) and retrospective (top) linear mixed effects models for 3 selected biomarkers.}
	\label{fig:resid_qq}     
\end{figure*}
\end{supplement}

\bibliographystyle{imsart-nameyear} 
\bibliography{covid_prediction.bib}       

\end{document}